\documentclass[english]{article}
\usepackage[T1]{fontenc}
\usepackage[latin9]{inputenc}
\usepackage{color}
\usepackage{amssymb,amsmath}
\usepackage{graphicx}


\usepackage{parskip}
\usepackage{url}
\usepackage{babel}
\usepackage[numbers,sort&compress]{natbib}
\begin{document}

\newtheorem{theorem}{Theorem}[section]
\newtheorem{lemma}[theorem]{Lemma}
\newtheorem{proposition}[theorem]{Proposition}
\newtheorem{corollary}[theorem]{Corollary}

\newenvironment{definition}[1][Definition]{\begin{trivlist}
\item[\hskip \labelsep {\bfseries #1}]}{\end{trivlist}}
\newenvironment{example}[1][Example]{\begin{trivlist}
\item[\hskip \labelsep {\bfseries #1}]}{\end{trivlist}}
\newenvironment{remark}[1][Remark]{\begin{trivlist}
\item[\hskip \labelsep {\bfseries #1}]}{\end{trivlist}}
\newcommand{\vectornorm}[1]{\left|\left|#1\right|\right|}

\newcommand{\rev}[1]{#1}

\title{Clustering memes in social media streams}
\author{Mohsen JafariAsbagh\and
		Emilio Ferrara\and
		Onur Varol\and
		Filippo Menczer\and
		Alessandro Flammini\bigskip\and
		School of Informatics and Computing.\\Indiana University. Bloomington IN (USA)}
\date{}%Received: date / Accepted: date}
\maketitle

\begin{abstract}
The problem of clustering content in social media has pervasive applications, including the identification of discussion topics, event detection, and content recommendation.
Here we describe a streaming framework for online detection and clustering of memes in social media, specifically Twitter. \rev{A pre-clustering procedure, namely protomeme detection, first isolates atomic tokens of information carried by the tweets.} Protomemes are thereafter aggregated, based on multiple similarity measures, to obtain memes as cohesive groups of tweets reflecting actual concepts or topics of discussion. 
The clustering algorithm takes into account various dimensions of the data and metadata, including natural language, the social network, and the patterns of information diffusion. As a result, our system can build clusters of semantically, structurally, and topically related tweets. 
The clustering process is based on a variant of Online K-means that incorporates a memory mechanism, used to ``forget'' old memes and replace them over time with the new ones.
The evaluation of our framework is carried out by using a dataset of Twitter trending topics. Over a one-week period, we systematically determined whether our algorithm was able to recover the trending hashtags. We show that the proposed method outperforms baseline algorithms that only use content features, as well as a state-of-the-art event detection method that assumes full knowledge of the underlying follower network.
\rev{We finally show that our online learning framework is flexible, due to its independence of the adopted clustering algorithm, and best suited to work in a streaming scenario.}
\end{abstract}

\section{Introduction}
Social media have become one of the main platforms for communication and information sharing. Popularity, high message exchange rate, social network structure, and accessibility across different devices make Twitter, in particular, ideal for spreading news \citep{bakshy2011everyone,kwak2010twitter}, discussing politics and social issues  \citep{conover2013geospatial,conover2013digital,varol2014evolution,morales2012users,skoric2011online}, and reaching out to groups or individuals with similar interests \citep{wu2011says}. On the other hand, social media are also used to disseminate misleading information such as rumors and spam \citep{chew2010pandemics,metaxas2010obscurity,ferrara2014rise}.
The ability to automatically detect, in real time, patterns of information diffusion or topics of conversation is of great interest. This task cannot be accomplished by classifying each single piece of content observed in real time on the platform; the classes cannot be predetermined because new memes constantly appear in the social stream, and labeling tweets for training is not feasible due to the large volume of content produced.

The alternative proposal pursued in this article is to design an unsupervised framework to cluster tweets based on similarity measures that account for content as well as other metadata features.
A single tweet may not contain sufficient information  to identify a topic of conversation, due to its brevity. Topic modeling techniques have proved ineffective in this task due to the sparsity of textual content \citep{hong2010empirical}. Metadata may help overcome this problem, as shown in other contexts \citep{mei2008topic}.
For example, in an attempt to mitigate the 140-character limitation, users may include URLs to provide external resources. Yet, most tweets don't contain URLs, and URLs may be abused to promote irrelevant content.
Another form of metadata is the \textit{mention}, a mechanism to include a specific user in a conversation. A mention can be either a direct reply to a user or a way of drawing her/his attention to the contents of the tweet. Mentions start with the \textit{@} character, followed by the user screen name. They can be exploited by spammers to grab someone's attention.
Users also enrich tweets with hashtags to identify topics of conversation.  Hashtags are user-defined terms starting with the \textsl{\#} character, and have been used as proxies for topics on Twitter \citep{tsur2012s,yang2012we}. Relying on hashtags alone as indicators of a topic has some limitations as well \citep{ferrara2013clustering}. For instance, they can be used to refer to a broad subject (e.g., \textsl{\#tcot} captures many conservative discussions on Twitter); or, two or more different hashtags can be used to refer to the same topic (e.g., \textsl{\#AffordableCareAct} and \textsl{\#ObamaCare} both talk about the \emph{Patient Protection and Affordable Care Act} of 2010).
Furthermore, users can adopt an organic hashtag to inject content into an ongoing conversation \citep{conover2011political}.\footnote{Example of hashtag injections include hijacked campaigns such as those by McDonald and the NYPD \citep{mcfail,nypd}.
%\#McFail? McDonald's Twitter Campaign Gets Hijacked: \url{http://www.cnbc.com/id/46132132}}$^,$
%NYPD Twitter campaign 'backfires' after hashtag hijacked: \url{http://www.bbc.com/news/technology-27126041}}
}

In our approach, we combine content and different sources of metadata to infer semantic similarity between groups of tweets.
We adopt the term \textit{meme} to denote sets of topically coherent tweets. A meme represents an idea or a concept that can spread from person to person in the social network. The goal of our unsupervised learning framework is to meaningfully cluster tweets from a social stream into potentially overlapping sets, each of which corresponds to a meme.

We leverage the various forms of metadata entities present in each tweet along with its content to bootstrap the clustering process. Specifically, we pre-aggregate tweets into initial sets based on shared metadata entities (URLs, mentions, and hashtags) and phrases --- sequences of word in the tweet text, excluding metadata entities. In the remainder of the paper we refer to these initial sets as \textit{protomemes}.

Our approach is based on the assumption that a protomeme represents a suitable atomic unit for the clustering algorithm, namely a semantically homogeneous set of tweets. On the other hand, there is a many-to-many relationship between tweets and protomemes: each tweet can contain more than one entity. As a result, protomemes don't commit the algorithm to assigning a tweet to a single cluster. From a practical point of view, a protomeme allows to build a textual representation that is less sparse than that of a single tweet. Protomemes yield higher accuracy in reconstructing political rumors in a static clustering scenario~\citep{ferrara2013clustering}. 

Although protomemes consist of a group of related tweets, they are an oversimplified representation of a meme, and therefore they cannot often capture all nuances of a topic. Protomemes can be too specific and only capture a particular aspect of a conversation, whereas a meme can have a more complex structure and may aggregate multiple perspectives into a single concept. Therefore, we assume that putting together protomemes in a meaningful way may yield sets of tweets which represent sophisticated memes better than any individual protomeme. In other words, our proposed approach benefits from protomemes as a pre-aggregation step and then it clusters them to identify memes.

Traditional clustering algorithms have been devised to perform in an offline scenario or stationary/static data. In such a case, the algorithm assumes unlimited access to all data points as well as boundless memory and processing resources. This scenario is far from reality: in the case of social media data, for example, posts arrive at varying rates and in a streaming fashion. In this situation, there is only a limited amount of time and resources available to process newly arrived data before they swamp the whole system. To operate in such an environment, it is essential to bear in mind that we can only afford one pass over the data or a few passes over a subset of the data. Problems of this sort have been studied in the area of data stream mining \citep{babcock2002models,gaber2005mining}, and they fall in the broader field of online learning \citep{albers1999online,blum1998online,cesa2006prediction,fiat1998online}.
Data stream clustering algorithms process a sequence of data points arriving over time.  Various such algorithms have been proposed in the literature \citep{gama2007learning,gama2010knowledge,sayed2012learning,shalev2011online}. Online K-means \citep{banerjee2004frequency,zhong2005efficient} is one well-known algorithm suitable for this purpose. While running, the algorithm maintains K clusters and it assigns each incoming data point to the closest cluster based on the distance between the new point and the centroid of the cluster. Here, we tailor this algorithm to take into account the notion of protomemes and our proposed similarity measures. This will allow us to design a system for clustering a social stream of content into memes.

\subsubsection*{Contributions and Outline}

In this paper, we propose a framework to extract content- and network-based features from Twitter data and use them to cluster the stream of tweets. A summary of the present contributions follows:

\begin{itemize}

\item \rev{We use the notion of protomemes \citep{ferrara2013clustering} as the result of our pre-aggregation step to group together tweets sharing some common information token.} We test the hypothesis that protomemes provide the clustering algorithm with useful atomic units.

\item We define various similarity measures between protomemes, based on content, metadata, and network features of Twitter data. We also introduce different ways to combine these similarity measures.

\item We design a stream clustering framework able to group protomemes together in topically cohesive clusters. Our method requires only one pass on the data, therefore is suitable to work with different online learning algorithms. We implement it by using a variant of Online K-means that works with protomemes and incorporates a sliding window mechanism.

\item We evaluate the performance of the proposed system on a stream of tweets. We show that our method outperforms two baseline algorithms, one solely based on tweet content and one based on tweet content plus the underlying social network structure. Our results demonstrate the advantage of adopting protomemes and various metadata features for meme clustering in social stream.

\end{itemize}

\section{Online Clustering Framework}
In the following we propose a general framework for clustering memes in social media data streams.
Our system can be applied to any social media system that supports directed relationships among users, and that generates a stream of messages produced by them. 
Twitter, Google Plus, and Tumblr are examples of such platforms. Hence, we will take the former as use case and we will adopt Twitter terminology to describe our platform.
Twitter users can engage in directed social relationships. 
One user can follow others (\textit{followees}), and, in turn, be followed by her/his own \textit{followers}.
Following is commonly used to receive other users' posts (\textit{tweets}) in one's news feed. 
Each user can also address others directly, by \textit{mentioning} their screen names in her/his tweets. 
Finally, users can re-broadcast (\textit{retweet}) messages from other users to make them visible by their followers.

Next, let us first define the notion of protomeme as the basic building block of our clustering framework. We then introduce various similarity measures between protomemes, and explain the data model that incorporates protomemes and allows us to design our meme stream clustering framework.

\subsection{Protomemes}
\rev{We recently proposed the notion of \emph{protomemes} as a simple way of grouping single messages into bigger blocks \cite{ratkiewicz2011truthy,ferrara2013clustering}. }
Our assumption is that these blocks yield topically cohesive groups and therefore represent natural units that can be aggregated into broader memes. 
To identify protomemes we simply group together tweets that share one instance of entities from the following types:

\begin{description}

\item[Hashtags:] Twitter users can incorporate in the text of their tweets one or more \emph{hashtags}, textual tokens prefixed by hash marks (\#), which identify the topic of the message. In a sense, hashtags leverage the \emph{wisdom of the crowd} \citep{golder2006usage,mika2007ontologies}: broad topics of interest emerge from the tagging activity of many users.

\item[Mentions:] We say that a tweet \emph{mentions} a user when it includes the target user's screen name preceded by the `@' symbol, thus addressing that specific user.

\item[URLs:] Tweets may include links to external sources of information. A \emph{URL} is the Web address identifying a linked resource. 

\item[Phrases:] We define the textual content of a tweet that remains after removing hashtags, mentions, URLs, stop words, and punctuation, and after stemming words~\citep{Porter80} as a \emph{phrase}. Phrases may help capture semantically equivalent variations of textual messages.

\end{description}

To demonstrate these different protomeme types, consider the following example: ``\textit{Tell your friends: \#Obamacare is helping young people afford health insurance. (via @OFATruthTeam) pic.twitter.com/s9QHilsSjO}.''

This tweet contains the hashtag \texttt{\#Obamacare}, the mention \texttt{@OFATruthTeam}, the URL \texttt{pic.twitter.com/s9QHilsSjO}, and the phrase \texttt{Tell friends help young people afford health insurance}. Each of these elements represents a different protomeme entity. All tweets containing the same entity will be grouped together. This way of defining protomemes also allows a single tweet to belong to multiple protomemes; the example above belongs to four protomemes. 
In the remainder of the paper we will interchangeably use the term \emph{protomeme} to refer to a set of tweets sharing the same entity, or to the entity itself.

In practice, the pre-aggregation step to extract protomemes requires only the application of a regular expression to the tweets in the data stream, which can be accomplished in real time \citep{ratkiewicz2011truthy}. 
Therefore, protomeme extraction can be achieved efficiently in a streaming scenario while requiring only a single pass on the data.

Next, we define various similarity measures between protomemes to further aggregate tweets and identify broader memes.

\subsection{Metadata and Similarity Measures}

\begin{figure}
\begin{center}
  \includegraphics[width=\textwidth]{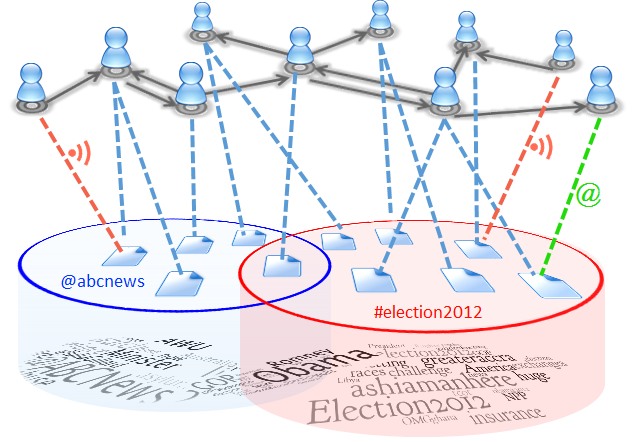}
  \caption{Relations among protomemes, tweets, users, and tweet content. There is a many-to-many relationship between memes and tweets. A user may be connected to a tweet as its author, by being mentioned in the tweet, or from retweeting the message.}
  \label{fig:protomeme_space}
\end{center}
\end{figure}

Fig.~\ref{fig:protomeme_space} illustrates the mutual relations between protomemes, the tweets they contain, their content, the users who post them, and the underlying follower network. 
We can define similarity measures between protomemes (or any sets of tweets) by considering the projections of the protomemes onto spaces induced by these features. 

Let us provide a few preliminary definitions. Let $p$ be a protomeme, and $\mathbf{p}^f$ a vector representation of $p$ according to feature $f$.
%Let $U_{\ell}$ be the set of users that produced tweets (and retweets)\footnote{Note that $|U_{\ell}| \leq |P_{\ell}|$ because a user may post more than one tweet.} in $P_{\ell}$. 
We can now  define a set of similarity measures based on content and metadata.

\begin{definition}[Common user similarity] $S^{u}$ between protomemes $p$ and $q$ is the cosine similarity  
\begin{equation}
S^u(p, q) = \frac{\mathbf{p}^u \cdot \mathbf{q}^u }{||\mathbf{p}^u|| ||\mathbf{q}^u||}
\label{eq:cus}
\end{equation}
where $\mathbf{p}^u$ and $\mathbf{q}^u$ are the vectors representing the number of tweets in $p$ and $q$, respectively, produced by each user.
\end{definition}

\begin{definition}[Content similarity] $S^c$ between $p$ and $q$ is the cosine similarity  
\begin{equation}
S^c(p, q) = \frac{\mathbf{p}^c \cdot \mathbf{q}^c }{||\mathbf{q}^c||  ||\mathbf{q}^c||}
\label{eq:cds}
\end{equation}
where $\mathbf{p}^c$ and $\mathbf{q}^c$ are the vectors representing the \emph{term frequency} (TF) weights assigned to each word of the two documents obtained by concatenating all the tweets in $p$ and $q$, respectively.\footnote{\rev{Term vectors might or might not include retweets; in all our experiments we include retweets. Our framework does not make any assumption on the language of the tweets either, therefore it is flexible to work with multiple languages.}}
\end{definition}

\begin{definition}[Common tweet similarity] $S^t$ between $p$ and $q$ is the cosine similarity 
\begin{equation}
S^t(p, q) = \frac{|p \cap q|}{\sqrt{|p||q|}}.
\label{eq:cts}
\end{equation}

\end{definition}

We wish to introduce a fourth similarity measure based on the social network of users involved in a set of tweets (see illustration of the follower network in Fig.~\ref{fig:protomeme_space}). Since we assume that information about the follower social network is not available to the clustering algorithm, let us exploit mention and retweet metadata as proxies for the underlying network and community structure \rev{whose role in information diffusion is crucial \citep{nematzadeh2014optimal}.} Let $N_p = U_p \cup M_p \cup R_p$ be the diffusion set of $p$, where $U_p$ is the set of authors of tweets in $p$, $M_p$ is the set of users mentioned in tweets in $p$, and $R_p$ is the set of authors of tweets with retweets in $p$.\footnote{Note that $R_p$ is not necessarily a subset of $U_p$ when only a sample of the tweets are considered in the stream; the sample may include a retweeted message but not the original one.} The diffusion set is a subset of the neighbors of users involved in the protomeme on the mention and retweet networks. 

\begin{definition}[Network similarity] $S^n$ between $p$ and $q$ is the cosine similarity between their diffusion sets
\begin{equation}
S^n(p, q) = \frac{|N_p \cap N_q|}{\sqrt{|N_p| |N_q|}}.
\label{eq:cns}
\end{equation}
\end{definition}

These four similarity measures can be combined to obtain a single similarity value between two protomemes. One common approach is to generate a \textit{linear combination} of similarity measures in the form of a weighted average \citep{sayyadi2009event,Qazvinian12}. Formally, we have 

\begin{equation}
S_{\mathcal{L}}(p, q) = \sum_k \omega_k S^k(p, q)
\label{eq:combination}
\end{equation}
with the constraint that $\sum_k \omega_k = 1$,  allowing for a normalized combination such that $S_{\mathcal{L}}(p, q) \in [0,1]$ (given that $ \forall k \; S^k \in [0,1]$).
The set of parameters $\omega_1, \dots, \omega_n$ yields an $n$-dimensional parameter space. Searching this space is necessary to identify the combination(s) of similarity measures that provide the optimal clustering performance. This is not a trivial task. 

A second approach is to choose the highest value among all similarity measures when computing the pairwise similarity between two protomemes. The rationale is that the relatedness of two protomemes may be best captured by different dimensions on a per-case base: for example, the similarity of a pair of protomemes might be best described by their content, while that of two other protomemes may be better reflected, for instance, by the social network dimension. Taking the maximum value of similarity among  the four dimensions would account for such heterogeneity. 
Given a set of similarity measures $S^1, \dots, S^n$, the \emph{maximum pairwise similarity} is formally defined as 
\begin{equation}
S_{\mbox{\tiny max}}(p, q) = \max_k \{S^k(p, q)\}.
\label{eq:maximization}
\end{equation}

The maximum pairwise similarity has an advantage over the linear combination, namely is that it does not need any parameter space exploration. 
More importantly, we found in previous work \citep{ferrara2013clustering} that the maximum pairwise similarity provides performance as good as the best linear combination in meme clustering. 
Hence, our next experiments will be based on maximum pairwise similarity.

\subsection{Data Processing Model}

The model assumption in data stream clustering is that, due to the large amount of incoming data, the system cannot store all of it in memory \citep{babcock2002models,gaber2005mining}.
Additionally, a data stream evolves with time and patterns in recent data are more  relevant for the clustering algorithm than those in older data. An established way to de-emphasize older data is to represent the stream trough a window-based model. 
There are three well-known window models in the literature: landmark window, damped window, and sliding window \citep{pramod2012data}. In all cases, without loss of generality, we assume the stream to start at time $t=0$ and discretize time into steps of fixed duration $\Delta t$.

A \emph{landmark window} contains all data points from $t=0$ until the present time $t=T$. This model is not feasible for fast-growing data streams.

A \emph{damped window} model assigns a weight to each data point in the data stream. It uses a decay function based on time, which gives more weight to recent data points. The most commonly used function attributes to an event occurring at time $t$ an exponentially decreasing weight $w(t)=2^{-\lambda (T-t)}$, where $\lambda>0$. The higher the value of $\lambda$, the higher the importance of more recent data.

A \emph{sliding window} has a fixed length of $\ell$ steps and at each moment $T$ it contains only the data points received during last $\ell$ steps, giving them equal importance. The window interval is $W = \left( T - \ell \Delta t, T \right]$. Algorithms that adopt this model either ignore data points older than $T - \ell \Delta t$, or consider a summary of them. The sliding window is a simple yet effective model in data stream processing. Due to its simplicity and generality, we choose this data processing model in our clustering algorithm. 

\subsection{Protomeme Stream Clustering}

Online K-means \citep{banerjee2004frequency,zhong2005efficient} is a simple data stream clustering algorithm based on iterative K-means for stationary data. In general, Online K-means starts with $K$ randomly chosen initial cluster seeds and every new point arriving in the stream is assigned to the closest existing cluster. The closest cluster is chosen based on the distance between the arriving point and the centroid of the cluster. A cluster centroid has the same features as the data points and the value of each feature is averaged across the data points that are members of the cluster.

This general algorithm does not take into account the fact that new concepts might appear in the stream, which are different from what has been observed before. These new concepts should represent new clusters; assigning them to the existing clusters might jeopardize the quality of clustering. 
To overcome this problem, one suggested approach \citep{aggarwal2012event} is to check whether the distance from the closest cluster centroid is an outlier in comparison to the other closest distances that have been observed so far. If not, the new data point is added to the nearest cluster. Otherwise, the least recently updated cluster is replaced by a new cluster with the new point as the only member. The least recently updated cluster is the one to which no new points have been assigned for the longest time. The outlier detection function uses a history of closest distances from previously observed data points to determine whether a given distance is an outlier. Every time a data point arrives in the stream, its distance to the closest centroid is added to the list. This method assumes that the distances follow a normal distribution. If the new distance exceeds the historical average by $n$ standard deviations or more, where $n$ is a parameter, the new point is deemed an outlier.  

The proposed clustering algorithm, that we call \emph{Protomeme Stream Clustering} (PSC), works as follows:

\begin{enumerate}

\item At the beginning of each step, the sliding window is advanced by $\Delta t$ and protomemes are extracted from arriving tweets in the stream, i.e., those in the new time step $(T-\Delta t, T]$. Each protomeme is treated as a data point to be clustered. Before any of these new points are assigned, all clusters are examined and data points with time stamps older than $T - \ell \Delta t$ (i.e., those that are no longer in the sliding window) are removed. From now on, we will refer to these points as \emph{old.} If a cluster consists only of old points, it becomes empty and is removed from the list of clusters.

\item Since we are using protomemes as a pre-aggregation step, in our algorithm we tend to assign the same protomemes to the same clusters whenever possible. If an arriving protomeme matches any of the ones present in any of the existing clusters, we assign it to that cluster and continue to the next protomeme. Otherwise, we move to the next step. 

\item A new protomeme is assigned to the closest cluster or to a new cluster based on the outcome of the outlier test. The protomeme is assigned to a new cluster if its distance from the nearest centroid $d > \mu + n \sigma$, where $\mu$ and $\sigma$ are the mean and standard deviation, respectively, of the values in the historical list of closest distance values. The historical distance values in the list are kept since the beginning of the clustering process.

\end{enumerate}

\begin{figure}
\begin{center}
  \includegraphics[width=\textwidth]{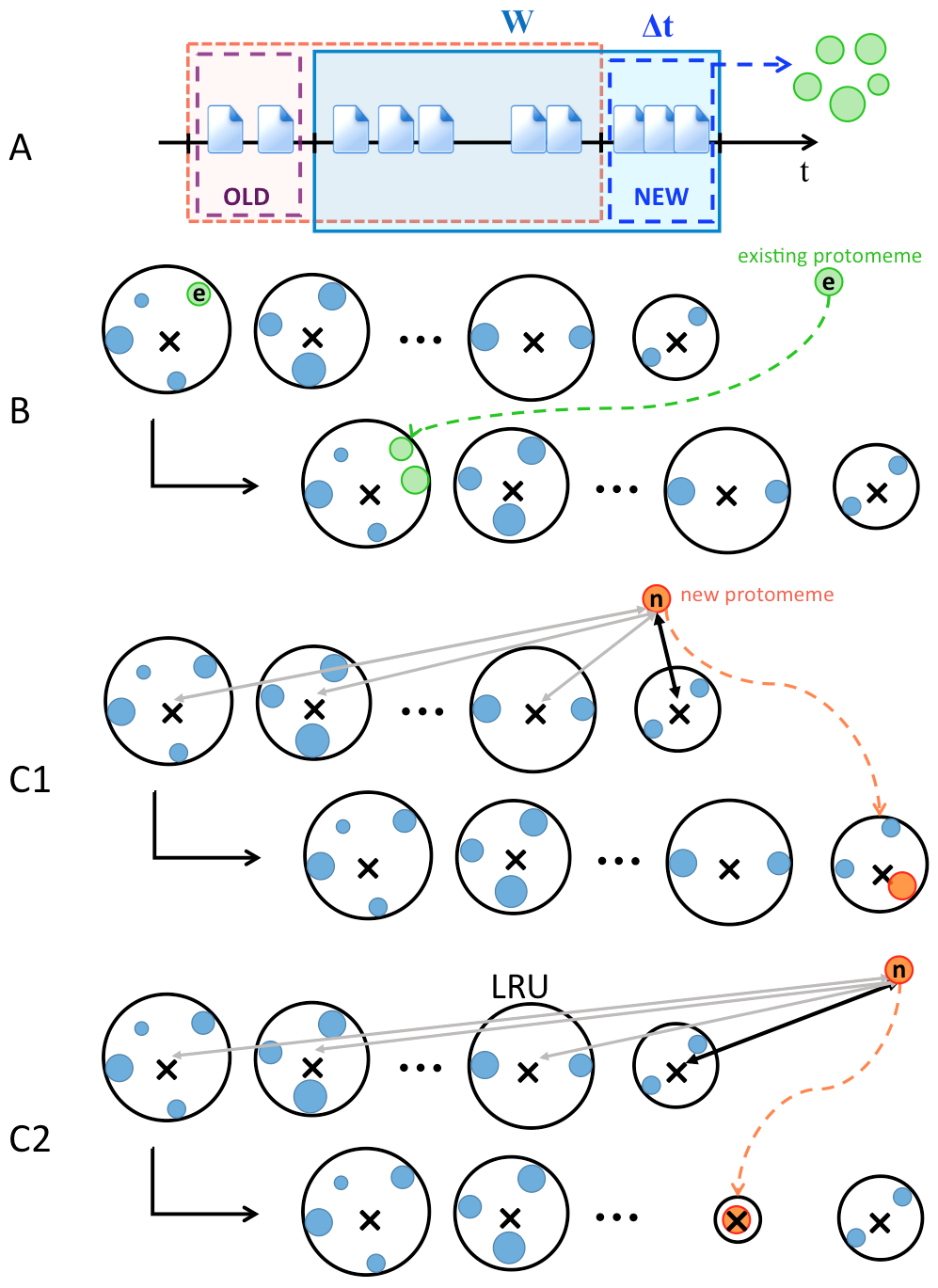}
  \caption{PSC stream clustering algorithm. (A) The window is shifted, old protomemes are removed, and new protomemes are extracted. (B) If an incoming protomeme exists, it is assigned to the same cluster. (C1 and C2) If an incoming protomeme is new, it is assigned to the closest cluster if it is not an outlier (C1), otherwise it replaces the least recently updated (LRU) cluster (C2).}
  \label{fig:clustering_algorithm}
\end{center}
\end{figure}

All the steps of the PSC clustering algorithm are depicted in Fig.~\ref{fig:clustering_algorithm}.

\section{Evaluation}

In this section we present the experimental procedure carried out to evaluate the performance of our clustering algorithm on a real dataset. First we describe the dataset on which we performed the experiments. Then we introduce the metric used to assess the quality of clustering.

\subsection{Ground Truth Dataset}

\begin{figure}
%\centering
%\begin{tabular}{lll|lll}
%\hline 
%Date & Hashtags & \multicolumn{1}{l}{Tweets} & Date & Hashtags & Tweets\tabularnewline
%\hline 
%2013-03-23 & $18$ & $3,322$ & 2013-03-27  & $13$ & $1,175$\tabularnewline
%2013-03-24 & $15$ & $1,449$ & 2013-03-28 & $11$ & $2,094$\tabularnewline
%2013-03-25 & $13$ & $678$ & 2013-03-29 & $10$ & $1,729$\tabularnewline
%\cline{4-6} 
%2013-03-26 & $13$ & $915$ & \multicolumn{1}{r}{\textbf{Total:}} & $27$ & $11,362$\tabularnewline
%\hline 
%\end{tabular}
\centerline{\includegraphics[width=\textwidth]{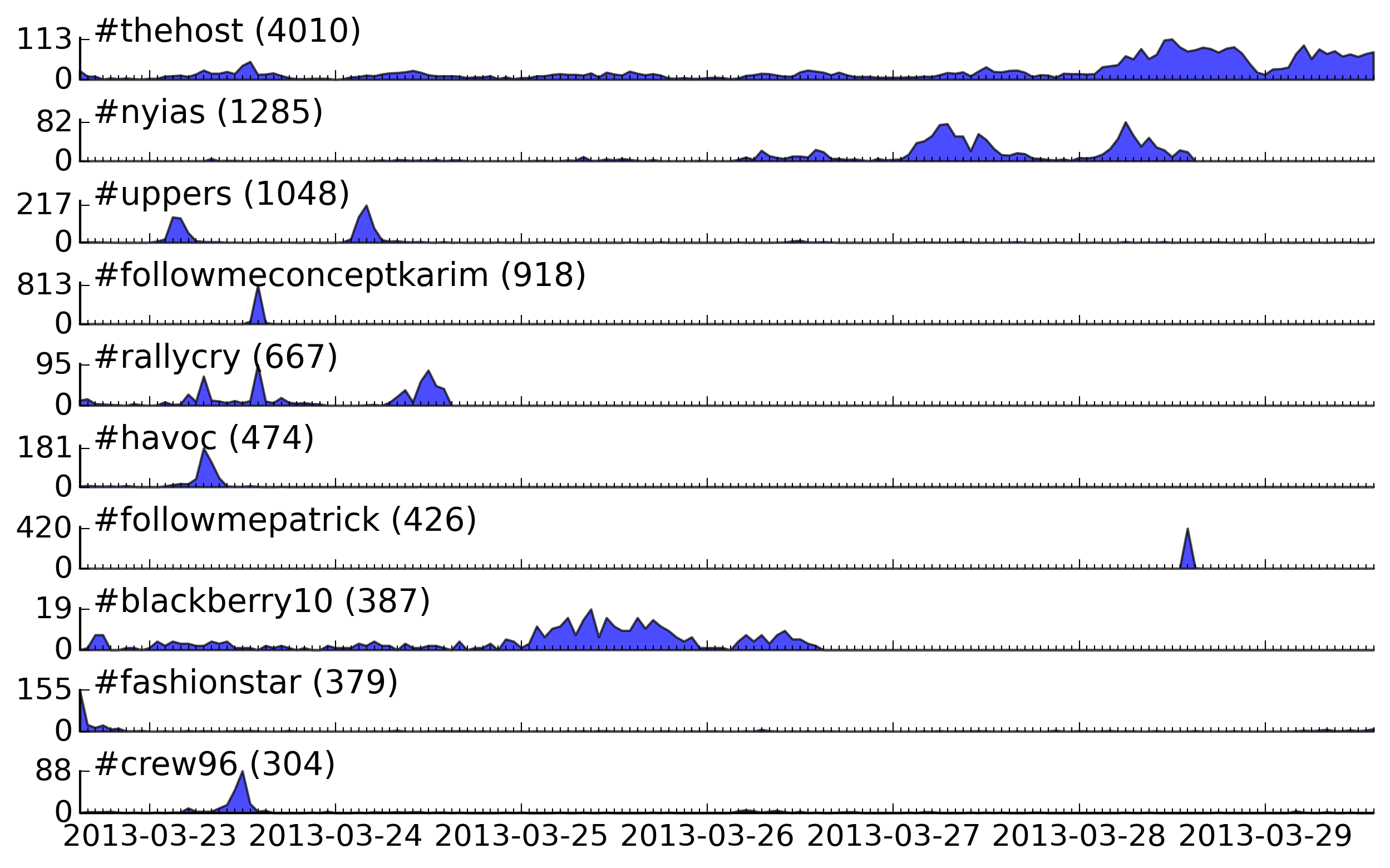}}
\caption {Tweet volume time series for the top 10 hashtags in the trending dataset. For each trending hashtag we report the total number of tweets with that hashtag in our 10\% Twitter sample (in parenthesis). Each bin is one hour.} 
\label{tab:dataset} 
\end{figure}

For the purpose of evaluation, we use a dataset collected from Twitter, comprised of trending hashtags between 23 and 29 March 2013. The data was obtained by monitoring and recording the trends appearing on the Twitter platform at regular intervals of 10 minutes. A detailed analysis of geographic and temporal trend diffusion based on this data is reported in prior work \citep{ferrara2013traveling}. We extracted all tweets containing those trending hashtags from a 10\% sample of the full stream of public tweets, for a time interval of 7 days before and 3 days after the trending point. The trending hashtags represent the ground truth memes --- we assume that tweets sharing a trending hashtag should be clustered together, even by an algorithm that does not have any knowledge of this cluster label. Figure~\ref{tab:dataset} shows a stream graph of the volume of tweets per hour, for the 10 most popular trending hashtags in our dataset. 

\subsection{Evaluation Metric}

To assess the quality of algorithmic cluster assignments, we adopt a measure based on \emph{Normalized Mutual Information} (NMI) \citep{danon2005comparing}. The NMI assumes the availability of a {\em ground truth} that represents the correct clusters. Let $A$ be the correct cluster assignment, and suppose that it contains $c_A$ clusters. Let $B$ be the output of a clustering algorithm operating on the same data and producing $c_B$ clusters. We can define a $c_A \times c_B$ \emph{confusion matrix} $\mathbf{N}$, whose rows correspond to the clusters in $A$ and whose columns represent clusters in $B$. Each entry $N_{ij}$ of this confusion matrix reports the number of elements of the correct $i$-th cluster that also happen to be present in the $j$-th cluster found by the clustering algorithm. Formally, the \emph{Normalized Mutual Information} is defined as
%{\small\[
\begin{equation} 
\mbox{NMI}(A,B) = \frac{-2 \displaystyle\sum_{i = 1}^{c_A}\sum_{j = 1}^{c_B} N_{ij} \log \left(\frac{N_{ij}N}{N_{i\cdot}N_{\cdot j}} \right)}{\displaystyle\sum_{i=1}^{c_A}N_{i \cdot} \log \left(\frac{N_{i \cdot}}{N}\right) + \displaystyle\sum_{j=1}^{c_B}N_{\cdot j} \log \left(\frac{N_{\cdot j}}{N}\right)}
\label{eq:nmi}
%\]
\end{equation}
%}
where $N_{i \cdot}$ (resp., $N_{\cdot j}$) is the sum of the elements in the $i$-th row (resp., $j$-th column) of the confusion matrix, and $N$ is the sum of all elements of $\mathbf{N}$. The output of this measure is normalized between zero (when the clusters in the two solutions are totally independent), and one (when they exactly coincide). Therefore, the higher the value of NMI, the better the quality of the clusters found by the algorithm.

Measures based on mutual information have been proved to best capture different facets of a clustering process, such as how well a clustering algorithm reflects the number, size, and composition of clusters with respect to the ground truth, as opposed to traditional information retrieval and data mining measures, like accuracy or purity, which usually produce biased evaluations \citep{meilua2007comparing}. Our investigation with accuracy, precision, recall, and $F_1$ confirmed the limitations of these measures, all of which report indistinguishable results due to the dominance of true negatives. Purity, on the other hand, is by definition biased toward rewarding the presence of tiny clusters, which tend to be pure. For these reasons, NMI has been recently adopted in the evaluation of tasks such as event detection in social media~\citep{becker2010learning}. 

Due to the fact that our algorithm can produce overlapping clusters of protomemes, we adopt a variant of NMI, called LFK-NMI after its authors \citep{lancichinetti2009detecting}, that is best suited to measure the quality of overlapping clusters thanks to a slightly different normalization criterion. In the remainder of the paper, we shall refer to LFK-NMI as NMI for simplicity.

\section{Results}

Let us now discuss the details of the PSC algorithm implementation and the values of the parameters for its configuration. We also introduce two baseline clustering algorithms against which to compare PSC.
%Then, we will present the results illustrating the comparison between our algorithm and two baseline algorithms.

\subsection{Experiment Setup}

Our experiments aim at assessing whether protomemes and our metadata-driven features and similarity measures bring an observable advantage in the task of clustering memes from social streams. To this end, we compare our framework against two baseline clustering algorithms that are also based on Online K-means, but operate directly on individual tweets and with different features and similarities. The description of the two baselines follows. 

\begin{description}

\item[Baseline B1:] 
This configuration is an implementation of the Online K-means clustering of simple tweets along with outlier handling as explained earlier. The only feature used in this algorithm is text content. The Term Frequency (TF) vector of each tweet is used to compute the content similarity between tweets and aggregate them.

\item[Baseline B2:] 
This configuration is an implementation of the event detection system recently proposed by Aggarwal and Subbian \citep{aggarwal2012event}, which is a tweet clustering algorithm based on a combination of content and network features. To the best of our knowledge, this approach represents the current state of the art in streaming clustering of tweets. It relies on the full knowledge of the follower network of all users present in the dataset. Such information provides a very significant advantage, but it also creates a practical challenge in that it is very time-consuming to obtain, making the algorithm infeasible in real-time, streaming scenarios. To compute tweet similarity, the original algorithm adopts TF-IDF, but we use TF in our implementation as it provides better performance on our dataset. This algorithm is also based on Online K-means and incorporates the same outlier handling procedure. To make use of this algorithm for comparison, we extracted in batch the follower network of all users present in our dataset.

\end{description}

As described in previous sections, all our Online K-means algorithms (PSC and baselines) have four parameters:

\begin{itemize}
\item[{$K$:}] initial number of clusters,
\item[{$\Delta t$:}] time step by which the window advances, 
\item[{$\ell$:}] length of sliding window, in time steps,
\item[{$n$:}] number of standard deviations from the mean to identify outliers. 
\end{itemize}

For all the algorithms, we set $\Delta t = 60$ minutes, $\ell = 6$ steps, and $n=2$. Therefore a sliding window has duration $\ell \Delta t = 360$ minutes. To set $K$ for online K-means, we computed the average number of hashtags across all sliding windows, yielding $K=11$ initial clusters. To evaluate the clustering solutions, we treat the trending hashtags as the actual (ground truth) cluster labels. Therefore, while extracting protomemes as data points to be clustered, we remove these trending hashtags from the set of protomemes, and from the text of the tweets as well. This prevents any bias in favor of our algorithm, and makes the clustering task very challenging. The evaluation score is computed at the end of each window, to which we will refer as evaluation period.

\subsection{Performance Evaluation}

\begin{figure}
\centerline{\includegraphics[width=1.0\linewidth]{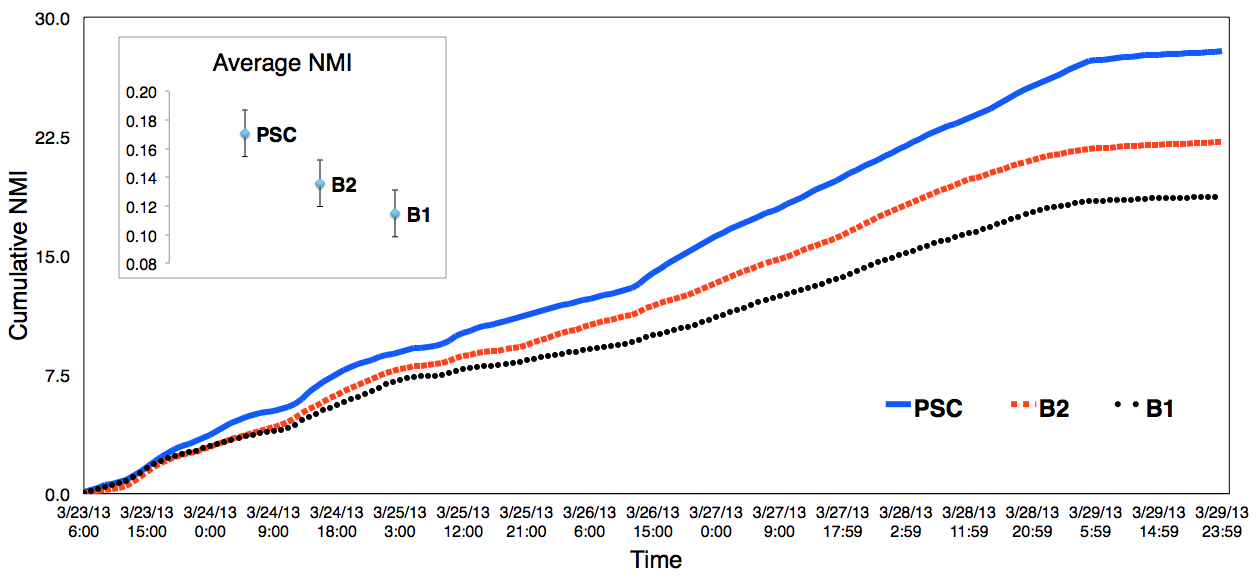}}
\caption{Performance of different clustering algorithms as a function of the evaluation period. For each algorithm, the NMI values at each step are averaged across five runs. These values are then accumulated over the course of the experiment. The inset plots the time-averaged NMI, with error bars corresponding to $\pm 1$ standard error. 
%For the sake of readability, the error bars are shown for selected points instead of all.
}
\label{fig:nmi_plot}
\end{figure}

Fig.~\ref{fig:nmi_plot} plots cumulative NMI over all the evaluation periods. Each point on the  \textit{x} axis represents a six-hour sliding window terminating at the indicated hour. 
To compute NMI correctly for each evaluation period, it is essential to have the same set of tweets in the ground truth and evaluated clusters. Therefore, we only use tweets and their labels in the ground truth for the same period of time. As explained earlier, whenever a cluster becomes empty after removing old data points, we remove it from the list of clusters. In a real world scenario, we might decide to ignore these clusters because they have not been updated during the last $\ell$ time steps; for evaluation purposes we keep them in a separate list and account for them when assessing the quality in the present window, then delete them afterwards.

Our algorithm performs consistently better than the two baselines. The performance improvement is more apparent after the online clustering has been carried out for an extended time. Fig.~\ref{fig:nmi_plot} shows that after about half of the running time, PSC provides a consistent improvement in cluster quality with respect to the baselines. This is due to the characteristic fast-paced churning time of the topics of discussion in social media. 
%We can also observe day-night patterns affecting the quality of the stream clustering uniformly across all methods. One possible explanation is the fact that the content production rate during days and nights is very different, therefore the algorithms require more time to adapt the clusters to newly observed topics. To get a better quantitative insight of how well our clustering algorithm performs, we computed the area under the ROC curve (AUROC) for the three algorithms. For each method, we show the box-plot reporting the average AUROC obtained across five runs of each algorithm. The results are presented in Fig.~\ref{fig:boxplot}. 
The inset of Fig.~\ref{fig:nmi_plot} demonstrates that the differences in NMI among PSC and the baseline algorithms are statistically significant. On average PSC outperforms B1 and B2 by 49\% and 26\%, respectively.

%\begin{figure}
%\begin{center}
%  \includegraphics[width=.8\linewidth]{boxplot}
%  \caption{Box-plot of the Area Under the ROC curve for five runs of each algorithm.}
%  \label{fig:boxplot}
%\end{center}
%\end{figure}

NMI is a quantitative measure that captures the overlap between the algorithmic \emph{clusters} and the \emph{classes} in the ground truth. It reports a single-number summary, but it does not provide any details about the resemblance between clusters and classes in terms of their numbers and size. For instance, if there is a huge class in the ground truth along with several small ones, an algorithm can achieve high NMI by assigning all the data points to a single cluster. 

\begin{figure}
\begin{center}
  \includegraphics[width=\textwidth]{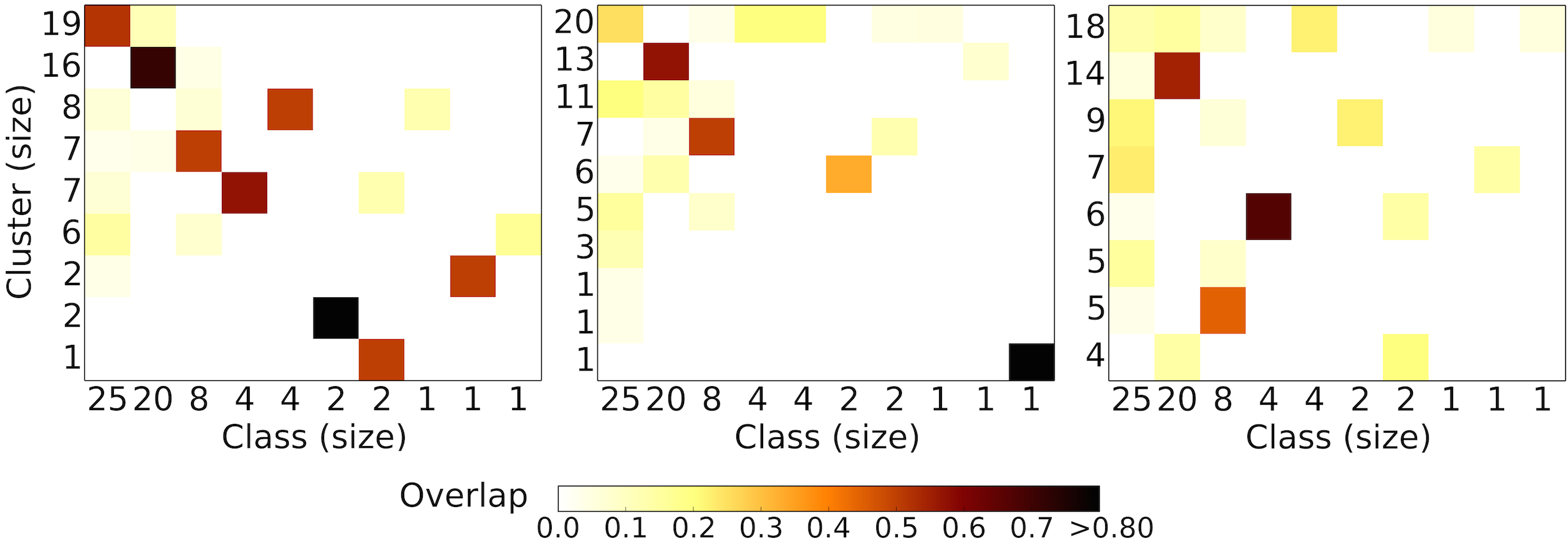}
\end{center}
\caption{Overlap (Jaccard coefficient) between ground truth classes and clusters detected by PSC (left), B2 (middle), and B1 (right). }
\label{fig:heatmap}
\end{figure}

To investigate the performance in greater detail, let us consider the confusion matrix containing the Jaccard coefficient between the set of tweets of every cluster in the solution and in the ground truth, respectively. Fig.~\ref{fig:heatmap} shows the confusion matrices for the three algorithms. The rows and columns in these matrices represent the clusters in the solution and classes in the ground truth, respectively. The number next to each row (resp., column) shows the number of tweets in each cluster (resp., class). These matrices are computed at an evaluation period 
%is marked by a dashed vertical line in Fig.~\ref{fig:nmi_plot}. The reason for this choice is that 
in which all three algorithms display local maxima in NMI. Although this period does not represent the best quality for any of the algorithms, it has the advantage that the ground truth classes are the same for all three algorithms, which is crucial for performance comparison.

A good clustering solution will have a confusion matrix with a dark colored cell (high value of Jaccard Coefficient) in each row or column. The perfect clustering would show only dark cells on the diagonal of a square confusion matrix. As Fig.~\ref{fig:heatmap} illustrates, PSC does a good job at capturing the actual clusters in the data; we observe greater confusion in the clusters generated by the two baseline algorithms.
In particular, our method is able to recover 8 clusters whose overlap with the ground truth cluster is above 60\%, while both the baseline methods identify at most 3 clusters faithfully resembling the ground truth.
Although the performance of the clustering methods fluctuates over time, 
%for reasons discussed above including the fast-paced churn time of Twitter discussion and the circadian cycles of content production, 
PSC is able to outperform the state of the art and discover memes in a streaming scenario with reasonable accuracy.
\rev{Our method also introduces great benefits in terms of computational cost: PSC processes streaming data points online, and its cost is linear in terms of the size of the input, similarly to the classic Online K-means  \cite{gaber2005mining}. Other methods require quadratic time and memory (for example hierarchical clustering \cite{ferrara2013clustering}), or access and storage of external information (e.g., B2 relies on the availability of the full follower network \cite{aggarwal2012event}).
}

\rev{\subsection{Parameter Tuning}

The stream clustering methods we presented here are all based on the setting of two parameters: the sliding window length $\ell$ and the step size $\Delta t$. We now investigate how the choice of such parameters affects the overall clustering performance. Let us explore the parameter space defined over the two dimensions of window length and step size. Figure \ref{fig:param_nmi} shows three heatmaps that illustrate how the clustering quality of each algorithm, measured by NMI, is affected by varying $\ell$ (represented on the y-axis) and $\Delta t$ (on the x-axis). All three algorithms achieve the best performance with the shortest time window ($\ell = \mbox{4 hours}$) and the smallest step size ($\Delta t = \mbox{30 min}$). This is somehow intuitive: the heterogeneity of data points (and, therefore, memes) collected with shorter windows is lower, and smaller steps allow to react more responsively to the fast-paced churning time of memes emerging on social platforms. 
With this parameter configuration, PSC outperforms both baselines, obtaining an average NMI score of 0.23, an increment of 43.75\% over B1 and 15\% over B2. 

As the window length and/or step size grow, the performance of all clustering methods worsens. However, the performance deterioration of PSC is very limited if compared with B1 and (especially) B2: the difference between best and worst PSC performance is 34.78\% (from an average NMI of 0.23 down to 0.15), whereas for B1 the decrease amounts for $43.75$\% (from 0.16 down to 0.09) and for B2 is 80\% (from 0.20 down to 0.04). This analysis shows that PSC is a more robust and flexible solution for stream clustering in social media, as it depends less on the optimal tuning of window size and step. Even if the computational resources are scarce, requiring coarser granularity, PSC displays reasonable accuracy and performs better than existing state-of-the-art methods.

%% FIL: it would be much better to use the same grayscale for all three plots!
\begin{figure}
\begin{center}
  \includegraphics[width=.33\textwidth, clip=true, trim=0 0 75 0]{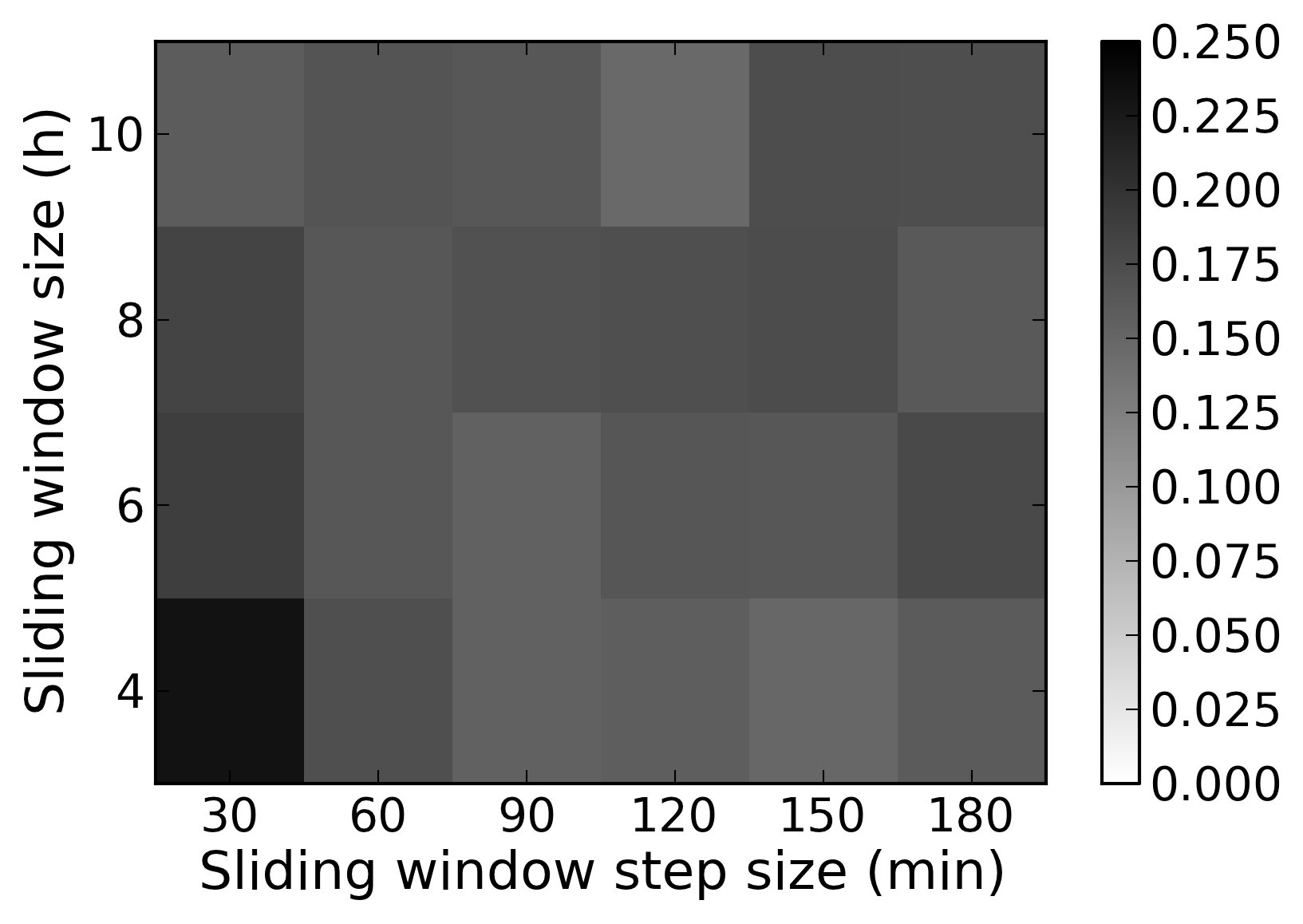}
  \includegraphics[width=.283\textwidth, clip=true, trim=45 0 75 0]{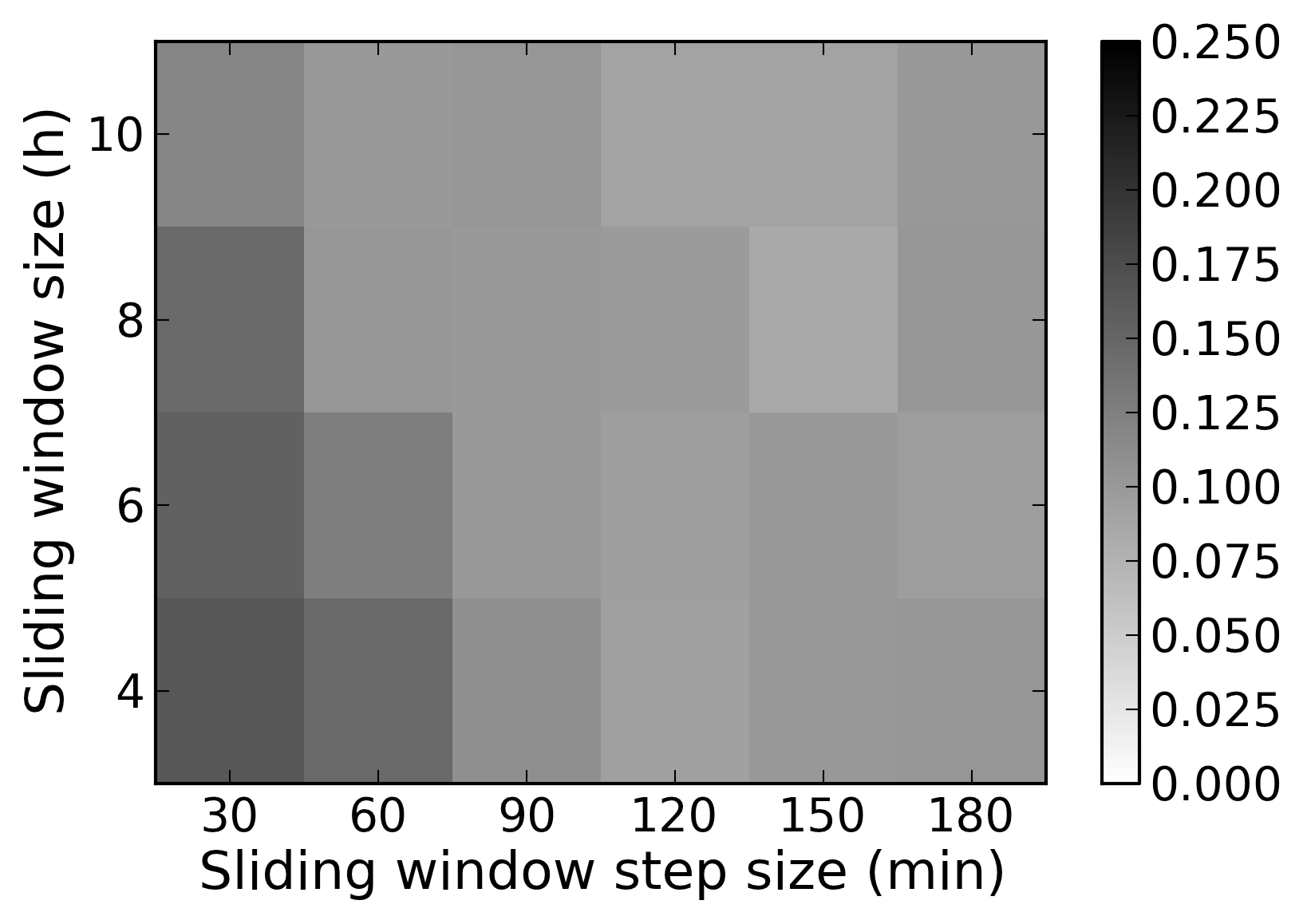}
  \includegraphics[width=.36\textwidth, clip=true, trim=45 0 0 0]{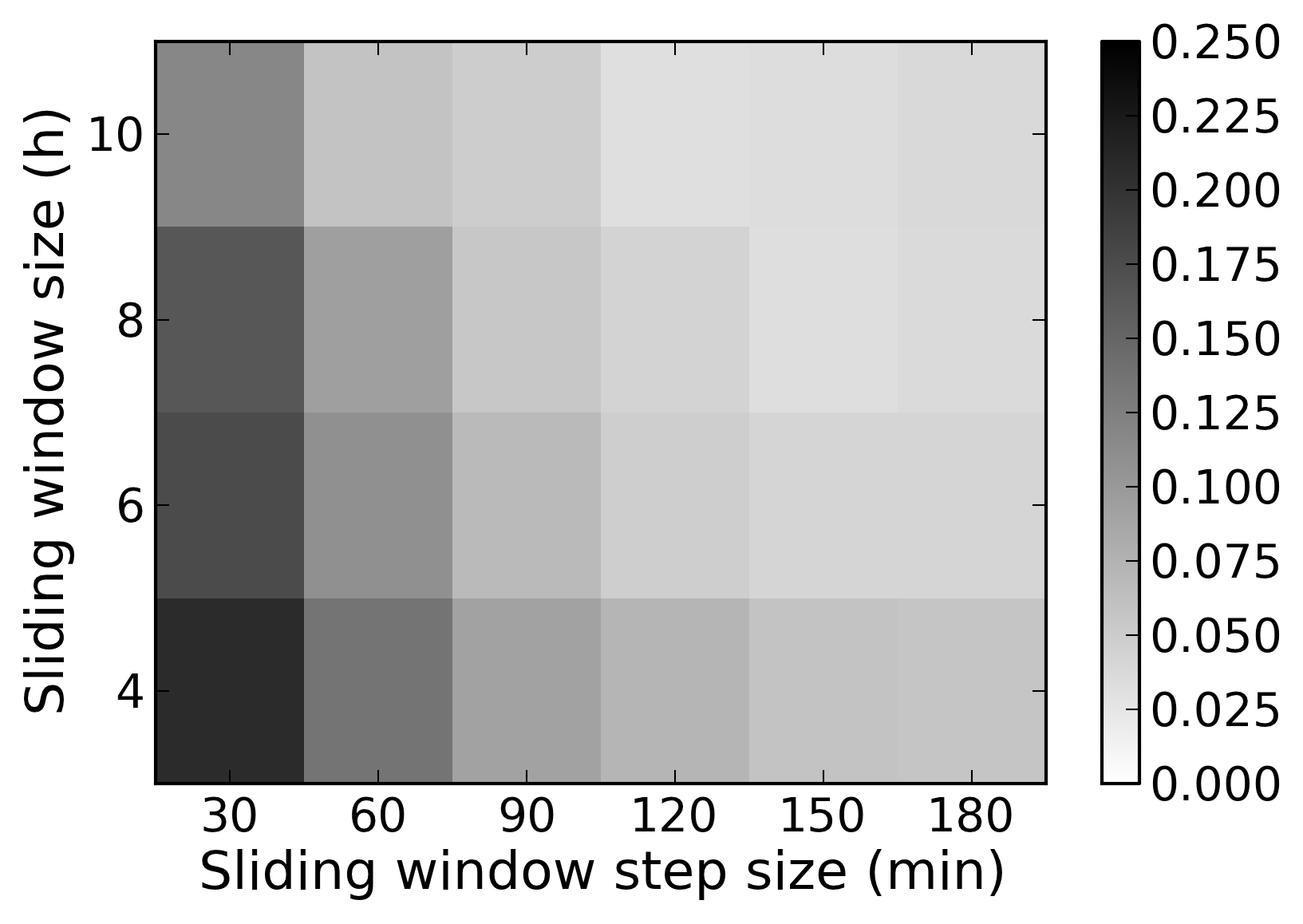}
\end{center}
\caption{\rev{Performance (measured by NMI) for varying sliding window length $\ell$ and step size $\Delta t$ of PSC (left), B1 (center) and B2 (right).}}
\label{fig:param_nmi}
\end{figure}

%\begin{figure}
%\begin{center}
%  \includegraphics[width=.32\textwidth]{OKM_parameterspace_purity}
%  \includegraphics[width=.32\textwidth]{OKM_TWB_parameterspace_purity}
%  \includegraphics[width=.32\textwidth]{AGG_parameterspace_purity}
%\end{center}
%\caption{Performance (measured by Purity) for varying sliding window lengths $\ell$ and steps $\Delta t$ of PSC (left), B1 (center) and B2 (right).}
%\label{fig:param_purity}
%\end{figure}

\begin{figure}
\begin{center}
  \includegraphics[width=\textwidth]{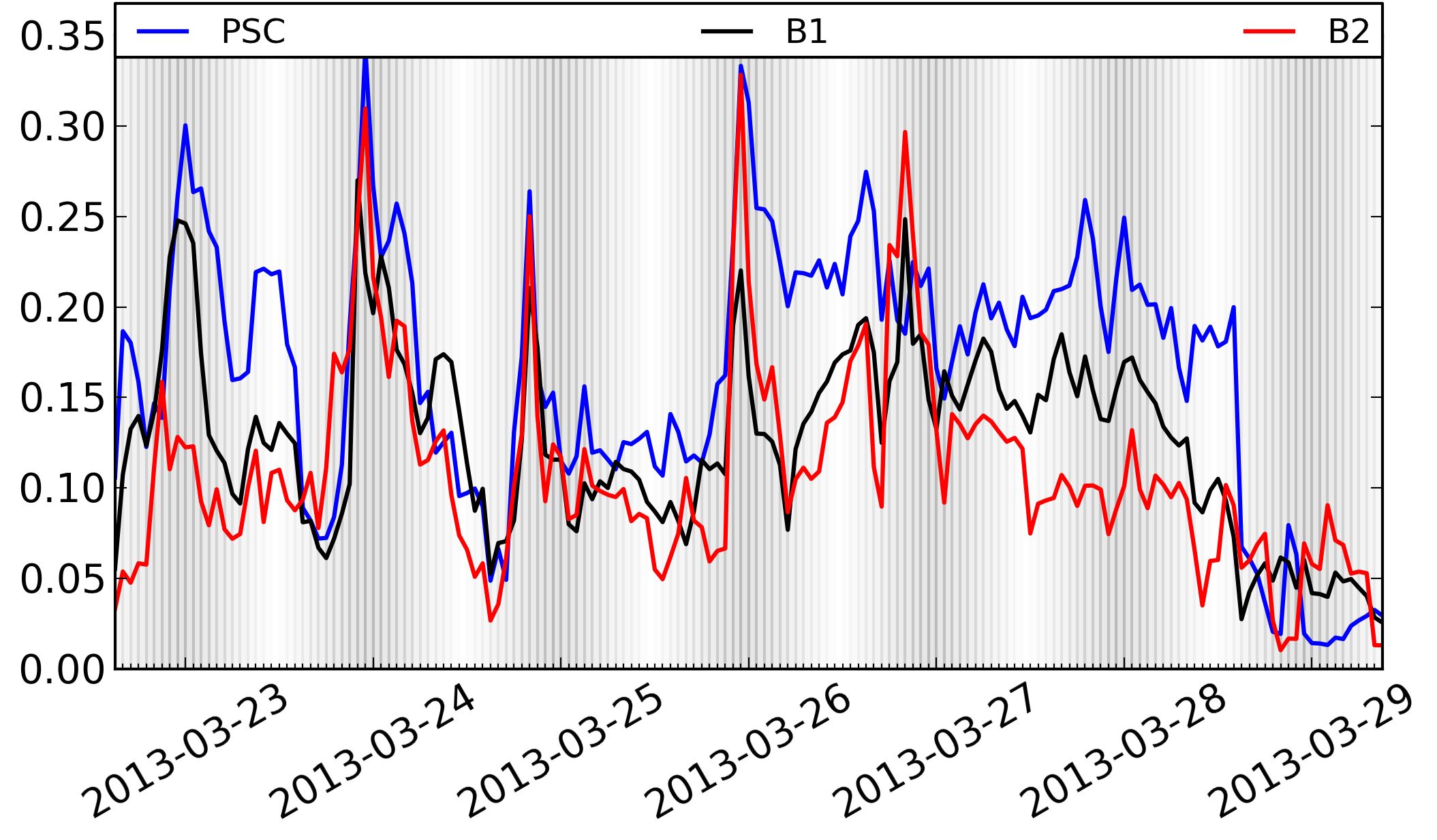}
\end{center}
\caption{\rev{Clustering quality over time (measured by NMI) achieved by PSC, B1 and B2. The gray bands in the background show the day-night cycle, with midnight in white.}}
%% FIL: is it correct that white = midnight?
\label{fig:overtime}
\end{figure}

\subsection{Success and Fail Scenarios}

Understanding when stream clustering performs well and when it fails is a challenging task. Here we try to identify what dynamics affect the outcome of the clustering process. We first focus on the temporal dimension: previous work shows that circadian rhythms of day-night activity affect the patterns of content generation and the volume of tweets (and memes) observed in social media \cite{golder2011diurnal}. We hypothesize that different rates of content production might determine how effective the clustering is: Figure \ref{fig:overtime} illustrates the performance of the three clustering algorithms over time. The background gray bands show the day-night cycle over a week-long period. All algorithms exhibit fluctuations in performance corresponding with the circadian clock: highs correspond to peek hours, where more content is generated and, therefore, more information is available to describe memes and their clusters; lows correspond with nightly hours, when the production of content is slowed and the clustering is based on fewer data points. PSC is consistently the best performing algorithms, as it reaches the highest peaks in clustering accuracy and the mildest dips. Over time, PSC seems to build on the accumulation of information over time to produce better results, as evident from the second part of the time series in Figure \ref{fig:overtime}. The other methods don't exhibit any benefit as time unfolds.

Let us explore a few examples of successful and unsuccessful stream meme clustering. Our goal is to identify other dynamics that might affect the quality of the clusters at the level of the single meme. To determine the quality of clustering of a single meme over time, let us introduce a new measure, called \emph{maximum cluster ratio} (MCR), defined as follows:
\begin{equation}
MCR(p) = \frac{\max_{c\in C}(N_{c}(p))}{N(p)}
\label{eq:mcr}
\end{equation}
%
%where $N(p)$ is the total number of tweets exhibiting protomeme $p$, and $N_{c}(p)$ is the number of tweets with protomeme $p$ in cluster $c\in C$. This measure allows us to dynamically determine over time how well the clustering algorithm is capturing a given protomeme; the higher MCR, the better.
where $N(p)$ is the total number of tweets exhibiting trending hashtag $p$, and $N_{c}(p)$ is the number of tweets with hashtag $p$ in cluster $c\in C$. This measure allows us to determine how well the algorithm is capturing a target meme over time; the higher the MCR, the better.
%% FIL: The definition in Eq 8 talked about protomemes but I am changing it to trending hashtags because, even if the definition is more general, we only introduce it for target clusters/memes. And those trending hashtags are removed, so they do not exist as protomemes, right?

Figure \ref{fig:success_fail} shows two scenarios: on the left, we display two memes (\#nyias and \#rallycry) whose clustering is of consistently high quality (high MCR); on the right, two examples of less successful clustering (\#blackberry10 and \#thehost) exhibit lower values of MCR. From our per case analysis a few considerations emerge: PSC performs generally better with shorter-lived memes (like \#nyias), and with memes whose content is produced mostly during day-time hours, as opposed to memes with sustained audience (see \#thehost); moreover, when content is abundant the performance is steady and the clustering of single memes is consistent and high quality (see \#rallycry), while when the volume of tweets associated with a memes is smaller the clustering quality can fluctuate (for example, see \#blackberry10). Another interesting fact is that the algorithm performs generally better with organic and grassroots memes (like in the examples on the left) rather than with memes related to promoted content. This difference might be attributed to the crucially different characteristics that such memes exhibit in terms of content generation and diffusion \cite{ferrara2013traveling}.

\begin{figure}
\begin{center}
  \includegraphics[width=\textwidth, clip=true, trim=0 0 0 22]{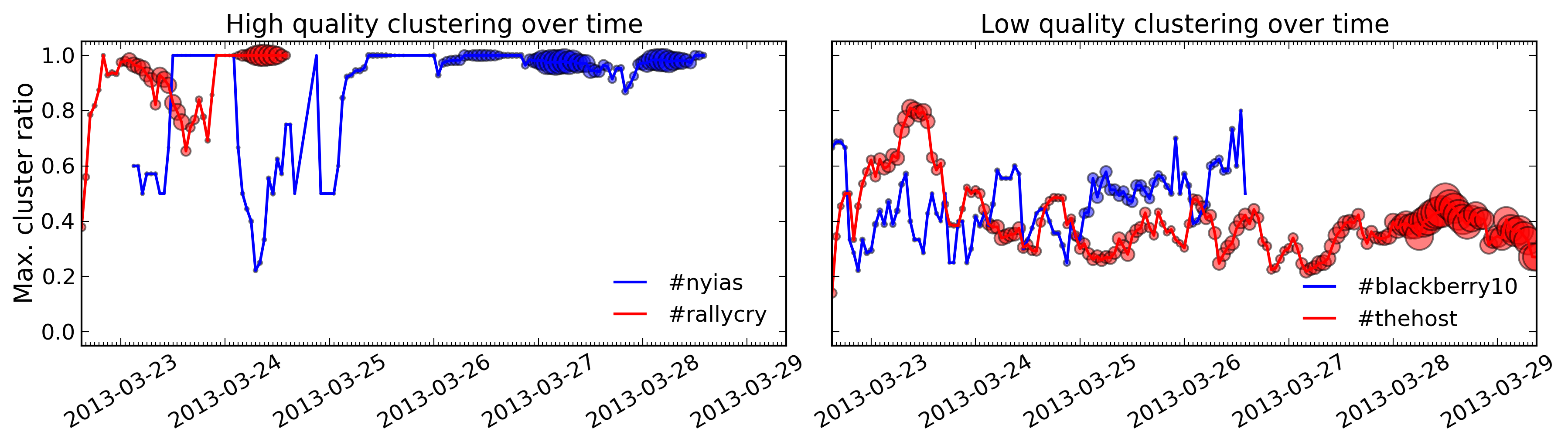}
\end{center}
\caption{\rev{Examples of two successful (left) and failed (right) attempts of capturing trending hashtag clusters over time with PSC. Clustering quality is measured by \emph{maximum cluster ratio}. The size of the circles is proportional to the total number of tweets with that hashtag at a given point in time.}}
\label{fig:success_fail}
\end{figure}

}

\section{Related Work} 	

Problems related to clustering in social media platforms include the identification of topics or memes \cite{leskovec2009meme,xie2011visual,simmons2011memes}, and event detection in social streams \cite{becker2011beyond,marcus2011twitinfo,lehmann2012dynamical,thom2012spatiotemporal,cataldi2013personalized}. 
Meme and topic identification techniques usually emphasize the terms and keywords as signatures of the content. 
%%%%% FIL begin: I do not understand the following. What is the difference between event detection and real-world event detection? It would seem one is a subset of the other, whereas the sentence below seems to suggest they are two different things.
Detection methods usually take into account temporal features and keywords occurrence to identify trending conversation, whereas to detect events happening in the physical world they mostly rely on temporal and geographical information. 
%%%%% FIL end
The present work, to the best of our knowledge, is the first to formalize the problem of clustering memes from online social media streams. 

Leskovec \textit{et al.} \cite{leskovec2009meme} presented \textsl{memetracker}, a platform for tracking memes originated in online media such as mainstream news sites and Weblogs. \textsl{Memetracker} can group together short, distinctive phrases that act as signatures of specific topics. The system can also identify small variations of them. It then creates groups of news on related topics that can be tracked over time to define patterns of diffusion in the news cycle. \textsl{Memetracker} assumes that the memes identified by aggregation on the basis of textual similarity are disjoint and correct: no systematic evaluation of the quality of the retrieved memes is provided. Our work instead focuses on the assessment of the quality of the meme clustering process, and allows for overlapping memes.

The problem of tracking news for meme extraction has been tackled also by Simmons \textit{et al.} \cite{simmons2011memes}. Based on the \textsl{memetracker} dataset,  the authors investigated whether information evolves and mutates while being consumed by social media users. Our definition of protomemes is rooted on the findings of both Leskovec \emph{et al.} \cite{leskovec2009meme} and Simmons \emph{et al.} \cite{simmons2011memes}, expanding on the aggregation of meme variants based not only on textual similarity, but also on other network and meta-data features.

Our framework shares some similarities also with another line of research on event detection systems. 
Aggarwal and Subbian \cite{aggarwal2012event} presented a clustering algorithm that exploits both content- and network-based features to detect events in social streams. We adapted their algorithm to work in the context of meme clustering, to use it as a baseline. Unfortunately, the algorithm assumes preexisting knowledge about the follower network of Twitter users. In a streaming scenario, such information is expensive to obtain, especially when encountering popular users. In our framework, we proposed to rely on mention and retweet diffusion sets, which can be inferred in real time from the observed data. 
Also, our system provides better performance by pre-clustering based on protomemes and by metadata-driven  measures of similarity between clusters. 

Becker \emph{et al.} \cite{becker2010learning} formulated the clustering problem for event detection on social media by defining a set of features and a supervised approach to combine them. The authors defined three types of features: textual (e.g., title, description, tags), time/date, and location. They proposed an ensemble clustering approach based on labeled data to combine the similarity revealed by each group of features. However, they did not consider the social network underlying social media platforms. The same authors presented an event classification system designed for Twitter \cite{becker2011beyond}. It incorporates a clustering module that exploits temporal, social, and topical features.

Finally, Thom \textit{et al.} \cite{thom2012spatiotemporal} developed a system for interactive analysis of location-based micro-blog messages, which can assist in the detection of real-world events in real time. This approach benefits from an online clustering algorithm based on X-means, a variation of K-means, to detect spatio-temporal dense topic clusters with a single term in common. The clustering algorithm extracts the terms from the messages and attaches a geolocation and a timestamp to each term. Then, comparison of Euclidean distance with a predefined threshold is used to assign each term to the closest cluster centroid or to a new cluster.

%Finally, Agarwal \emph{et al.} \cite{agarwal2012real} recently proposed a graph-based algorithm for the real-time discovery of clusters in dynamic networks. The strategy is based on the discovery of dense clusters ($\frac{1}{2}$-\emph{quasi} cliques) on the inferred graph of correlated keywords, extracted from tweets in a given time-frame. This method relies on the adoption of the \emph{short cycle property} that allows for a local approximate solution of an otherwise NP-hard task. Performance of the system has been benchmarked by using a simulated stream of tweets based on events reported by Google news in a given period, yielding high precision/recall in the task of identifying the top $k$ largest events. 

\section{Conclusions} 

In this work we proposed a framework to deal with the problem of clustering memes in social media streams, Twitter in particular. Our system is based on a pre-clustering procedure, called protomeme detection, aimed at identifying atomic tokens of information contained in each tweet. This strategy only requires text processing, therefore is particularly efficient and well suited for a streaming scenario. 
Memes are thereafter obtained by aggregating protomemes on the basis of the similarity among them, computed by  \textit{ad-hoc} measures defined according to various dimensions including content, the social network, and information diffusion patterns. Such measures only adopt information that can be extracted in a streaming fashion from observed data, and they allow to build clusters of topically related tweets. 
The meme clustering is carried out by using a variant of Online K-means, which integrates a memory mechanism to keep track of the least recently updated memes. 
We used a dataset comprised of trending hashtags on Twitter to systematically evaluate the performance of our algorithm and we showed that our method outperforms a baseline that only uses tweet text, as well as one that assumes full knowledge of the underlying social network.

\rev{
One crucial feature of our system is that it can be extended to work with any clustering algorithm based on similarity (or distances). In this paper, for example, we chose to present Online K-means because of its simplicity; however, during our design we also tested other methods including density-based and hierarchical data stream clustering algorithms (e.g., \textit{DenStream} \cite{cao06densitybasedstream} and \textit{LiarTree} \cite{kranen11hierarchicalstream}). Although a complete benchmark and tuning of these alternative methods was out of the scope of our analysis, we collected evidence of the ease of extension of our framework to different algorithms. }

In the future one could extend the set of features incorporated by our clustering framework, considering for instance entities such as images. Furthermore, our preliminary analysis suggests that the introduction of time series as features may yield significant performance improvements.
Our long-term plan is to integrate the meme clustering framework with a meme classifier to distinguish engineered types of social media communication from spontaneous ones. This platform will adopt supervised learning techniques to classify memes and determine their legitimacy, with the aim to detect misinformation and deception campaigns in their early stages. The platform will be optimized to work with the real-time, high-volume streams of messages typical of Twitter and other online social media.

\bibliographystyle{abbrv}
\bibliography{bibliography}

\begin{thebibliography}{10}

\bibitem{aggarwal2012event}
C.~Aggarwal and K.~Subbian.
\newblock Event detection in social streams.
\newblock In {\em Proceedings of SIAM International Conference on Data Mining},
  2012.

\bibitem{albers1999online}
S.~Albers and S.~Leonardi.
\newblock Online algorithms.
\newblock {\em ACM Computing surveys}, 31(3), 1999.

\bibitem{babcock2002models}
B.~Babcock, S.~Babu, M.~Datar, R.~Motwani, and J.~Widom.
\newblock Models and issues in data stream systems.
\newblock In {\em Proceedings of the twenty-first ACM SIGMOD-SIGACT-SIGART
  symposium on Principles of database systems}, pages 1--16. ACM, 2002.

\bibitem{bakshy2011everyone}
E.~Bakshy, J.~Hofman, W.~Mason, and D.~Watts.
\newblock Everyone's an influencer: quantifying influence on twitter.
\newblock In {\em Proceedings of the 4th ACM International Conference on Web
  Search and Data Mining}, pages 65--74. ACM, 2011.

\bibitem{banerjee2004frequency}
A.~Banerjee and J.~Ghosh.
\newblock Frequency-sensitive competitive learning for scalable balanced
  clustering on high-dimensional hyperspheres.
\newblock {\em Neural Networks, IEEE Transactions on}, 15(3):702--719, 2004.

\bibitem{nypd}
BBC.
\newblock {NYPD Twitter campaign 'backfires' after hashtag hijacked}.
\newblock \url{http://www.bbc.com/news/technology-27126041}, 2014.

\bibitem{becker2010learning}
H.~Becker, M.~Naaman, and L.~Gravano.
\newblock Learning similarity metrics for event identification in social media.
\newblock In {\em Proceedings of the 3rd ACM International Conference on Web
  Search and Data Mining}, pages 291--300. ACM, 2010.

\bibitem{becker2011beyond}
H.~Becker, M.~Naaman, and L.~Gravano.
\newblock Beyond trending topics: Real-world event identification on twitter.
\newblock In {\em Proceedings of the 5th International AAAI Conference on
  Weblogs and Social Media}, 2011.

\bibitem{blum1998online}
A.~Blum.
\newblock {\em On-line algorithms in machine learning}.
\newblock Springer, 1998.

\bibitem{cao06densitybasedstream}
F.~Cao, M.~Ester, W.~Qian, and A.~Zhou.
\newblock Density-based clustering over an evolving data stream with noise.
\newblock In {\em In 2006 SIAM Conference on Data Mining}, pages 328--339,
  2006.

\bibitem{cataldi2013personalized}
M.~Cataldi, L.~D. Caro, and C.~Schifanella.
\newblock Personalized emerging topic detection based on a term aging model.
\newblock {\em ACM Transactions on Intelligent Systems and Technology (TIST)},
  5(1):7, 2013.

\bibitem{cesa2006prediction}
N.~Cesa-Bianchi.
\newblock {\em Prediction, learning, and games}.
\newblock Cambridge University Press, 2006.

\bibitem{chew2010pandemics}
C.~Chew and G.~Eysenbach.
\newblock Pandemics in the age of twitter: content analysis of tweets during
  the 2009 h1n1 outbreak.
\newblock {\em PLoS One}, 5(11):e14118, 2010.

\bibitem{mcfail}
CNBC.
\newblock {\#McFail? McDonald's Twitter Campaign Gets Hijacked}.
\newblock \url{http://www.cnbc.com/id/46132132}, 2013.

\bibitem{conover2011political}
M.~Conover, J.~Ratkiewicz, M.~Francisco, B.~Gon{\c{c}}alves, F.~Menczer, and
  A.~Flammini.
\newblock Political polarization on twitter.
\newblock In {\em ICWSM}, 2011.

\bibitem{conover2013geospatial}
M.~D. Conover, C.~Davis, E.~Ferrara, K.~McKelvey, F.~Menczer, and A.~Flammini.
\newblock The geospatial characteristics of a social movement communication
  network.
\newblock {\em PloS one}, 8(3):e55957, 2013.

\bibitem{conover2013digital}
M.~D. Conover, E.~Ferrara, F.~Menczer, and A.~Flammini.
\newblock {The digital evolution of Occupy Wall Street}.
\newblock {\em PloS one}, 8(5):e64679, 2013.

\bibitem{danon2005comparing}
L.~Danon, A.~D{\'\i}az-Guilera, J.~Duch, and A.~Arenas.
\newblock Comparing community structure identification.
\newblock {\em Journal of Statistical Mechanics: Theory and Experiment},
  2005(09):P09008, 2005.

\bibitem{ferrara2013clustering}
E.~Ferrara, M.~JafariAsbagh, O.~Varol, V.~Qazvinian, F.~Menczer, and
  A.~Flammini.
\newblock Clustering memes in social media.
\newblock In {\em Proceedings of the 2013 IEEE/ACM International Conference on
  Advances in Social Networks Analysis and Mining}, pages 548--555. IEEE/ACM,
  2013.

\bibitem{ferrara2014rise}
E.~Ferrara, O.~Varol, C.~Davis, F.~Menczer, and A.~Flammini.
\newblock The rise of social bots.
\newblock {\em arXiv preprint arXiv:1407.5225}, 2014.

\bibitem{ferrara2013traveling}
E.~Ferrara, O.~Varol, F.~Menczer, and A.~Flammini.
\newblock Traveling trends: social butterflies or frequent fliers?
\newblock In {\em Proceedings of the first ACM conference on Online social
  networks}, pages 213--222. ACM, 2013.

\bibitem{fiat1998online}
A.~Fiat and G.~Woeginger.
\newblock {\em Online algorithms: The state of the art}.
\newblock Springer Heidelberg, 1998.

\bibitem{gaber2005mining}
M.~M. Gaber, A.~Zaslavsky, and S.~Krishnaswamy.
\newblock Mining data streams: a review.
\newblock {\em ACM Sigmod Record}, 34(2):18--26, 2005.

\bibitem{gama2007learning}
J.~Gama and M.~M. Gaber.
\newblock {\em Learning from data streams}.
\newblock Springer, 2007.

\bibitem{gama2010knowledge}
J.~Gama, P.~P. Rodrigues, E.~J. Spinosa, and A.~C. P. L.~F. de~Carvalho.
\newblock {\em Knowledge discovery from data streams}.
\newblock Chapman \& Hall/CRC Boca Raton, 2010.

\bibitem{golder2006usage}
S.~Golder and B.~Huberman.
\newblock Usage patterns of collaborative tagging systems.
\newblock {\em Journal of information science}, 32(2):198--208, 2006.

\bibitem{golder2011diurnal}
S.~A. Golder and M.~W. Macy.
\newblock Diurnal and seasonal mood vary with work, sleep, and daylength across
  diverse cultures.
\newblock {\em Science}, 333(6051):1878--1881, 2011.

\bibitem{hong2010empirical}
L.~Hong and B.~Davison.
\newblock Empirical study of topic modeling in twitter.
\newblock In {\em Proceedings of the 1st Workshop on Social Media Analytics},
  pages 80--88. ACM, 2010.

\bibitem{kranen11hierarchicalstream}
P.~Kranen, F.~Reidl, F.~S. Villaamil, and T.~Seidl.
\newblock Hierarchical clustering for real-time stream data with noise.
\newblock In {\em Proc. of the 23nd International Conference on Scientific and
  Statistical Database Management (SSDBM 2011), Portland, Oregon, USA}, pages
  405--413, Heidelberg, Germany, 2011. Springer.

\bibitem{kwak2010twitter}
H.~Kwak, C.~Lee, H.~Park, and S.~Moon.
\newblock What is twitter, a social network or a news media?
\newblock In {\em Proceedings of the 19th International Conference on World
  Wide Web}, pages 591--600. ACM, 2010.

\bibitem{lancichinetti2009detecting}
A.~Lancichinetti, S.~Fortunato, and J.~Kert{\'e}sz.
\newblock Detecting the overlapping and hierarchical community structure in
  complex networks.
\newblock {\em New Journal of Physics}, 11(3):033015, 2009.

\bibitem{lehmann2012dynamical}
J.~Lehmann, B.~Gon{\c{c}}alves, J.~Ramasco, and C.~Cattuto.
\newblock Dynamical classes of collective attention in twitter.
\newblock In {\em Proceedings of the 21st International Conference on World
  Wide Web}, pages 251--260, 2012.

\bibitem{leskovec2009meme}
J.~Leskovec, L.~Backstrom, and J.~Kleinberg.
\newblock Meme-tracking and the dynamics of the news cycle.
\newblock In {\em Proceedings of the 15th ACM SIGKDD International Conference
  on Knowledge Discovery and Data Mining}, pages 497--506. ACM, 2009.

\bibitem{marcus2011twitinfo}
A.~Marcus, M.~Bernstein, O.~Badar, D.~Karger, S.~Madden, and R.~Miller.
\newblock Twitinfo: aggregating and visualizing microblogs for event
  exploration.
\newblock In {\em Proceedings of the 2011 Annual Conference on Human Factors in
  Computing Systems}, pages 227--236. ACM, 2011.

\bibitem{mei2008topic}
Q.~Mei, D.~Cai, D.~Zhang, and C.~Zhai.
\newblock Topic modeling with network regularization.
\newblock In {\em Proceedings of the 17th international conference on World
  Wide Web}, pages 101--110. ACM, 2008.

\bibitem{meilua2007comparing}
M.~Meil{\u{a}}.
\newblock Comparing clusterings -- an information based distance.
\newblock {\em Journal of Multivariate Analysis}, 98(5):873--895, 2007.

\bibitem{metaxas2010obscurity}
P.~Metaxas and E.~Mustafaraj.
\newblock From obscurity to prominence in minutes: Political speech and
  real-time search.
\newblock In {\em Proceedings of Web Science: Extending the Frontiers of
  Society On-Line}, 2010.

\bibitem{mika2007ontologies}
P.~Mika.
\newblock Ontologies are us: A unified model of social networks and semantics.
\newblock {\em Web Semantics: Science, Services and Agents on the World Wide
  Web}, 5(1):5--15, 2007.

\bibitem{morales2012users}
A.~Morales, J.~Losada, and R.~Benito.
\newblock Users structure and behavior on an online social network during a
  political protest.
\newblock {\em Physica A: Statistical Mechanics and its Applications}, 2012.

\bibitem{nematzadeh2014optimal}
A.~Nematzadeh, E.~Ferrara, A.~Flammini, and Y.-Y. Ahn.
\newblock Optimal network modularity for information diffusion.
\newblock {\em Physical Review Letters}, 113(8):088701, 2014.

\bibitem{Porter80}
M.~Porter.
\newblock An algorithm for suffix stripping.
\newblock {\em Program}, 14(3), 1980.

\bibitem{pramod2012data}
S.~Pramod and O.~Vyas.
\newblock Data stream mining: A review on windowing approach.
\newblock {\em Global Journal of Computer Science and Technology Software \&
  Data Engineering}, 12(11):26--30, 2012.

\bibitem{ratkiewicz2011truthy}
J.~Ratkiewicz, M.~Conover, M.~Meiss, B.~Gon{\c{c}}alves, S.~Patil, A.~Flammini,
  and F.~Menczer.
\newblock Truthy: mapping the spread of astroturf in microblog streams.
\newblock In {\em Proceedings of the 20th International Conference Companion on
  World Wide Web}, pages 249--252. ACM, 2011.

\bibitem{sayed2012learning}
M.~Sayed-Mouchaweh and E.~Lughofer.
\newblock {\em Learning in non-stationary environments}.
\newblock Springer, 2012.

\bibitem{sayyadi2009event}
H.~Sayyadi, M.~Hurst, and A.~Maykov.
\newblock Event detection and tracking in social streams.
\newblock In {\em Proceedings of the 3rd International AAAI Conference on
  Weblogs and Social Media}, 2009.

\bibitem{shalev2011online}
S.~Shalev-Shwartz.
\newblock Online learning and online convex optimization.
\newblock {\em Foundations and Trends in Machine Learning}, 4(2):107--194,
  2011.

\bibitem{simmons2011memes}
M.~Simmons, L.~A. Adamic, and E.~Adar.
\newblock Memes online: Extracted, subtracted, injected, and recollected.
\newblock In {\em Proceedings of the 5th International AAAI Conference on
  Weblogs and Social Media}. AAAI, 2011.

\bibitem{skoric2011online}
M.~Skoric, N.~Poor, Y.~Liao, and S.~Tang.
\newblock Online organization of an offline protest: From social to traditional
  media and back.
\newblock In {\em Proceedings of the 44th Hawaii International Conference on
  System Sciences}, 2011.

\bibitem{thom2012spatiotemporal}
D.~Thom, H.~Bosch, S.~Koch, M.~Worner, and T.~Ertl.
\newblock Spatiotemporal anomaly detection through visual analysis of
  geolocated twitter messages.
\newblock {\em Visualization Symposium, IEEE Pacific}, 0:41--48, 2012.

\bibitem{tsur2012s}
O.~Tsur and A.~Rappoport.
\newblock What's in a hashtag?: content based prediction of the spread of ideas
  in microblogging communities.
\newblock In {\em Proceedings of the fifth ACM international conference on Web
  search and data mining}, pages 643--652. ACM, 2012.

\bibitem{varol2014evolution}
O.~Varol, E.~Ferrara, C.~L. Ogan, F.~Menczer, and A.~Flammini.
\newblock Evolution of online user behavior during a social upheaval.
\newblock In {\em Proceedings of the 2014 ACM conference on Web science}, pages
  81--90. ACM, 2014.

\bibitem{wu2011says}
S.~Wu, J.~Hofman, W.~Mason, and D.~Watts.
\newblock Who says what to whom on twitter.
\newblock In {\em Proceedings of the 20th International Conference on World
  Wide Web}, pages 705--714. ACM, 2011.

\bibitem{xie2011visual}
L.~Xie, A.~Natsev, J.~R. Kender, M.~Hill, and J.~R. Smith.
\newblock Visual memes in social media: tracking real-world news in youtube
  videos.
\newblock In {\em Proceedings of the 19th ACM International Conference on
  Multimedia}, pages 53--62. ACM, 2011.

\bibitem{yang2012we}
L.~Yang, T.~Sun, M.~Zhang, and Q.~Mei.
\newblock We know what@ you\# tag: does the dual role affect hashtag adoption?
\newblock In {\em Proceedings of the 21st international conference on World
  Wide Web}, pages 261--270. ACM, 2012.

\bibitem{Qazvinian12}
W.~Yih and V.~Qazvinian.
\newblock Measuring word relatedness using heterogeneous vector space models.
\newblock In {\em Proceedings of Annual Conference of the North American
  Chapter of ACL}, 2012.

\bibitem{zhong2005efficient}
S.~Zhong.
\newblock Efficient online spherical k-means clustering.
\newblock In {\em Neural Networks, 2005. IJCNN'05. Proceedings. 2005 IEEE
  International Joint Conference on}, volume~5, pages 3180--3185. IEEE, 2005.

\end{thebibliography}

\end{document}